\pdfoutput=1
\documentclass[aps,prb,twocolumn,noshowpacs,a4paper,superscriptaddress]{revtex4-1}

\usepackage{graphicx}
\usepackage{color}
\usepackage{textcomp}
\usepackage{amssymb}\usepackage{amsmath}\usepackage{amsthm}
\usepackage{units}
\usepackage{bm}
\usepackage{bbold}
\usepackage{braket}

\begin{document}

\title{Antiferromagnetic correlations in two-dimensional fermionic Mott-insulating and metallic phases}

\author{J. H. Drewes}
\thanks{These authors contributed equally to this work.}
\affiliation{Physikalisches Institut, University of Bonn, Wegelerstrasse 8, 53115 Bonn, Germany}
\affiliation{Cavendish Laboratory, University of Cambridge, JJ Thomson Avenue, Cambridge CB3 0HE, United Kingdom}
\author{L. A. Miller}
\thanks{These authors contributed equally to this work.}
\affiliation{Physikalisches Institut, University of Bonn, Wegelerstrasse 8, 53115 Bonn, Germany}
\affiliation{Cavendish Laboratory, University of Cambridge, JJ Thomson Avenue, Cambridge CB3 0HE, United Kingdom}
\author{E. Cocchi}
\thanks{These authors contributed equally to this work.}
\affiliation{Physikalisches Institut, University of Bonn, Wegelerstrasse 8, 53115 Bonn, Germany}
\affiliation{Cavendish Laboratory, University of Cambridge, JJ Thomson Avenue, Cambridge CB3 0HE, United Kingdom}
\author{C. F. Chan}
\affiliation{Physikalisches Institut, University of Bonn, Wegelerstrasse 8, 53115 Bonn, Germany}
\author{D. Pertot}
\affiliation{Physikalisches Institut, University of Bonn, Wegelerstrasse 8, 53115 Bonn, Germany}
\author{F. Brennecke}
\affiliation{Physikalisches Institut, University of Bonn, Wegelerstrasse 8, 53115 Bonn, Germany}
\author{M. K{\"o}hl}
\email{michael.koehl@uni-bonn.de}
\affiliation{Physikalisches Institut, University of Bonn, Wegelerstrasse 8, 53115 Bonn, Germany}

\maketitle

\textbf{Near zero temperature, quantum magnetism can non-trivially arise from short-range interactions, but the occurrence of magnetic order depends crucially on the interplay of interactions, lattice geometry, dimensionality and doping. Even though the consequences of this interplay are not yet fully understood, quantum magnetism is believed to be connected to a range of complex phenomena in the solid state, for example, in the context of high-$T_c$ superconductivity\cite{Dagotto1994,Lee2006} and spin liquids in frustrated lattices\cite{Balents2010}. Ultracold atomic Fermi gases in optical lattices\cite{Esslinger2010} constitute an experimental system with unrivalled tunability and detection capabilities to explore quantum magnetism by analog quantum simulation. In this work, we study the emergence of antiferromagnetic (AFM) correlations between ultracold fermionic atoms in two dimensions with decreasing temperature. We determine the  magnetic susceptibility of the Hubbard model from simultaneous  measurements of the in-situ density of both spin components. At half-filling and strong interactions our data approach the Heisenberg model of localized spins with antiferromagnetic correlations. Moreover, we observe the disappearance of magnetic correlations when the system is doped away from half-filling. Our observation of the dependence of quantum magnetism on doping paves the way for future studies on the emergence of pseudogap and pairing phenomena away from half-filling.}

A spin-1/2 mixture of ultracold fermionic atoms in an optical lattice emulates electrons in a crystalline lattice of a solid\cite{Bloch2008,Esslinger2010}. Hence, it can be employed as a highly-controlled model system to study the phases of strongly correlated matter\cite{Hofstetter2002,Kohl2005b,Joerdens2008,Schneider2008}, which has remained a challenge both for experiment and theory. A paradigm for describing the physics of fermionic particles on a lattice is the Hubbard model\cite{Hubbard1963}, which describes the complex many-body problem by the fundamental processes of tunnelling between neighbouring lattice sites with amplitude $t$ and interactions of two particles with opposite spin on the same lattice site with energy $U$. The occurrence and properties of low-temperature quantum phases are governed by these two parameters.  At half-filling (i.e. on average one particle per lattice site), when the interaction strength dominates other energy scales $U \gg t, k_BT$, the model describes a crossover from a metal to a Mott insulator in which the density becomes ordered with exactly one fermion per lattice site\cite{Joerdens2008,Schneider2008,Taie2012,Duarte2015,Greif2016,Cocchi2016,Cheuk2016,Hofrichter2015,Drewes2016}. At lower temperatures $k_BT \sim J = 4t^2/U$, the system is able to further minimise its energy by developing non-local AFM correlations in the spin degree of freedom. The formation of magnetic correlations across the lattice from purely on-site  interactions in the absence of direct magnetic long-range interaction is one of the hallmarks of quantum magnetism. Whether or not these  magnetic correlations lead to long-range order depends crucially on the dimensionality: in three dimensions a second-order phase transition is expected to occur at finite temperature; in two dimensions at finite temperature magnetic order must remain of finite spatial extent due to enhanced quantum fluctuations. The interplay of dimensionality, quantum magnetism originating from superexchange, and its breakdown when doping the lattice with particles or holes, has been intensively, albeit inconclusively, investigated in the context of the pseudogap and high-temperature superconductivity in the cuprates\cite{Dagotto1994,Lee2006}.

\begin{figure*}
 \includegraphics[width=2\columnwidth,clip=true]{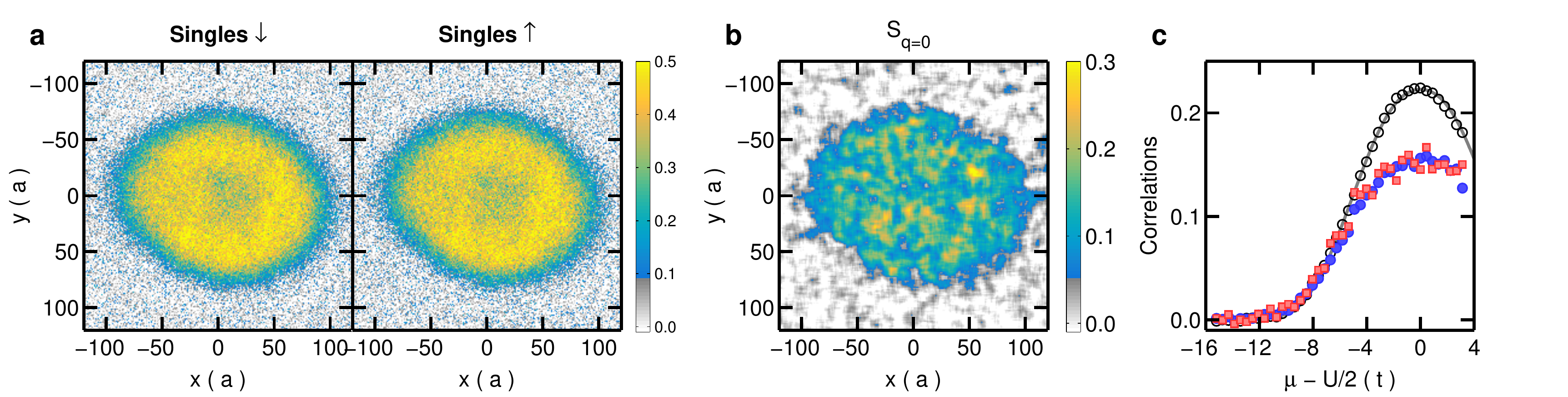}
 \caption{Spin-resolved singles density distributions and spin correlations. (a) In-situ distributions of both spin-up and spin-down densities on singly-occupied lattice sites averaged over 130 experimental realisations for $U/t=8.2(5)$ and $k_BT/t = 0.96(2)$. (b) Spatial map of the spin structure factor $S_{\bm{q}=0}$  extracted from the data set shown in (a) and spatially averaged over a region of $3 \times 3$ pixels for better visibility. (c) Local moment $C_{0,0}$ shown as open circles and structure factor $S_{\bm{q}=0}$ shown as red squares ($\xi  =\unit[4]{\mu m}$) and blue filled circles ($\xi  =\unit[1.7]{\mu m}$) as a function of chemical potential $\mu$ (see Methods). The solid line is a fit of the local moment to NLCE data\cite{Khatami2011}, serving as an independent thermometer.}
 \label{fig1}
\end{figure*}

The magnetic properties of a system are characterized by the magnetic susceptibility $\chi$, which is defined as the linear response of the magnetisation $M\simeq M_0+\chi H$ to a weak external, uniform magnetic field $H$.  Here $M_0$ is the magnetisation without any external field, which is zero in  our case. At high temperature, the microscopic magnetic moments are randomly oriented by thermal excitations and hence the magnetic  susceptibility approaches zero. At low temperature, magnetic correlations build up, yielding a finite value of $\chi$. In solids and liquids the magnetic susceptibility can be measured by nuclear magnetic resonance techniques. However, this  averages over macroscopically large volumes, which sometimes masks the relationship between the thermodynamic quantity $\chi$ and its microscopic origin.

Ultracold atomic Fermi gases in optical lattices are an ideal platform for studying quantum magnetism in order to reveal the microscopic origin of the magnetic susceptibility in the Hubbard model. Previous experiments have detected evidence for antiferromagnetic correlations in measurements averaging over inhomogeneous systems\cite{Hart2015,Greif2013} 
and, in parallel with the work presented here, spatial magnetic correlations have been detected using quantum gas microscopes\cite{Parsons2016,Boll2016,Cheuk2016a}. We detect the emergence of AFM correlations upon reducing temperature through the magnetic susceptibility and its dependence on temperature, interaction strength and filling. To this end, we employ high-resolution in-situ imaging\cite{Cocchi2016} of the density distributions of {\em both} spin components in {\em a single measurement} and determine the spin structure factor and the magnetic susceptibility. Since the magnetic correlations due to the superexchange interaction are strongest at half-filling and sensitively depend on the filling, we employ spatially-resolved detection and hence determine the spin structure factor as a function of doping. We compare our results to state-of-the-art calculations and find agreement where results are available, which is, in particular, at half filling. Away from half-filling theoretical data are sparse, since in two dimensions even the most advanced theoretical models are not able to probe the low-temperature regime realised in this work.

Our experiment begins with a spin-balanced quantum degenerate mixture of fermionic $^{40}K$ atoms prepared in the two lowest hyperfine states $|F=9/2,m_F=-9/2\rangle$ and $|F=9/2,m_F=-7/2\rangle$ denoted in the following as spin-up ($\uparrow$) and spin-down ($\downarrow$), respectively\cite{Cocchi2016}. We load the gas into an anisotropic, three-dimensional optical lattice with strongly suppressed tunnelling along the vertical direction and an approximately square lattice geometry along the horizontal directions. For a lattice depth of $6 E_\mathrm{rec}$ along the horizontal axes, where $E_\mathrm{rec} = \hbar^2 \pi^2/(2ma^2)$ and $a = 532$\,nm is the lattice constant, we realise a tunnelling amplitude of $t = h \times 224$\,Hz. We control the Hubbard interaction parameter over the range $1.6(2)\leq U/t \leq 12.0(7)$ by employing a magnetic Feshbach resonance at $\unit[202]{G}$ and tune the temperature of the gas in the lattice by periodically modulating the lattice depth at approximately twice the trapping frequency of the confining potential.

The observable of interest, $\hat{S}^z_i = (\hat{n}_{i,\uparrow}-\hat{n}_{i,\downarrow})/2=(\hat{s}_{i,\uparrow}-\hat{s}_{i,\downarrow})/2$, depends only on the spin-up and spin-down distributions of the singly-occupied sites (singles) where $\hat{n}_{i,\sigma}$ ($\hat{s}_{i,\sigma} = \hat{n}_{i,\sigma}-\hat{n}_{i,\uparrow} \hat{n}_{i,\downarrow}$) denote the number operator for particles (singles) in spin state $\sigma = \uparrow, \downarrow$ at lattice site $i$. Therefore, after freezing an equilibrium distribution in a deep horizontal lattice, we remove all atoms residing on doubly-occupied sites by employing exothermic spin-changing collisions (see Methods).  The  in-situ singles density distributions of both spin components within a single horizontal layer are detected by performing simultaneous radio-frequency tomography on both spin states in a vertical magnetic field gradient and subsequent high-resolution absorption imaging. We precisely calibrate the underlying trapping potential $V(\bm{r})$ resulting from the Gaussian envelope of the optical lattice beams and employ the local density approximation (LDA) to average over isopotentials and access correlations as a function of local chemical potential and filling\cite{Cocchi2016}.

\begin{figure*}
 \includegraphics[width=2\columnwidth,clip=true]{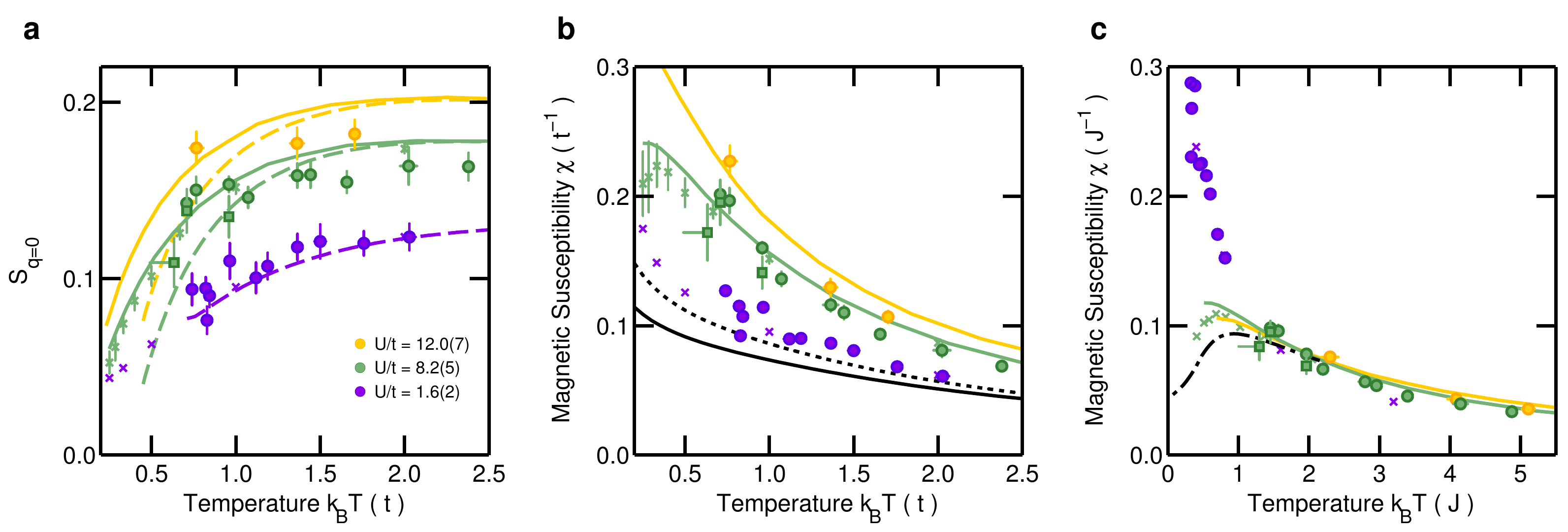}
 \caption{Antiferromagnetic correlations at half-filling. (a) Measured spin structure factor $S_{\bm{q}=0}$ for different interactions and temperatures. Squares correspond to $\xi  =\unit[4]{\mu m}$ and circles  to $\xi  =\unit[1.7]{\mu m}$ (see Figure 1c). NLCE data\cite{Khatami2011} for $S_{\bm{q}=0}$ (solid lines) and $|C_{0,0}|-4 |C_{0,1}|$ (dashed lines) are shown for $U/t=2$ (purple), $8$ (green) and $12$ (yellow). Crosses show QMC results\cite{Paiva2010} of $S_{\bm{q}=0}$ for $U/t=2$ (purple) and $8$ (green). (b) Magnetic susceptibility $\chi$ extracted from data shown in (a). Solid lines and crosses show corresponding results from NLCE and QMC. The solid black line shows the magnetic susceptibility $\chi$ for vanishing interactions and the dotted black line for weak interactions $U/t = 2$ in the random-phase approximation\cite{Hirsch1985}. (c) Magnetic susceptibility scaled in units of the superexchange interaction $J$ and comparison to the antiferromagnetic Heisenberg model\cite{Sandvik2010} (dashed-dotted line).}
 \label{fig2}
\end{figure*}

In order to quantify the magnitude of AFM correlations and extract the magnetic susceptibility we perform a correlation analysis on a set of pairs of in-situ images $\{\tilde{s}_\uparrow(\bm{r}),\tilde{s}_\downarrow(\bm{r})\}$ of the singles distributions taken under equal experimental conditions. In the following, quantities with a tilde refer to snapshots of local observables convolved with the imaging point spread function and projected onto the imaging frame with coordinate $\bm{r}$. We compute the spatial correlations of $\tilde{S}^z(\bm{r}) = (\tilde{s}_\uparrow(\bm{r})-\tilde{s}_\downarrow(\bm{r}))/2$ over the set of images and integrate them over a region of radius $\xi$ (see Methods). For $\xi$ being large compared to the expected correlation length and our imaging resolution of $\unit[1.2(1)]{\mu m}$, the integrated correlations equal the zero-momentum spin structure factor $S_{\bm{q}=0} = \sum_{j}\left(\langle \hat{S}^z_{i} \hat{S}^z_{j}\rangle-\langle \hat{S}^z_{i} \rangle \langle \hat{S}^z_{j}\rangle\right)$ where $\bm{q}$ denotes the wave vector\cite{Hung2011}. For AFM correlations, the sign of the spin-spin correlator $C_{i_x,i_y} = \langle \hat{S}^z_{0,0} \hat{S}^z_{i_x,i_y}\rangle -\langle \hat{S}^z_{0,0}\rangle \langle \hat{S}^z_{i_x,i_y}\rangle$ between lattice sites with difference vector $(i_x,i_y)a$ alternates according to $(-1)^{i_x+i_y}$. Therefore, the spin structure factor has alternating contributions from different distances, beginning with the local moment, $S_{\bm{q}=0}=|C_{0,0}|-4|C_{0,1}|+4|C_{1,1}|...$\,. At the temperatures realized in this work, the magnitudes of $C_{i_x,i_y}$ are expected to decay exponentially with distance\cite{LeBlanc2013}. According to the fluctuation-dissipation theorem\cite{Kubo1966}, the uniform magnetic susceptibility is given by $\chi = S_{\bm{q}=0}/(k_B T)$ where $k_B$ denotes the Boltzmann constant. To obtain the spin structure factor and the magnetic susceptibility as a function of chemical potential (resp.~filling) we average the magnetic correlations over regions of constant local chemical potential $\mu = \mu_0-V(\bm{r})$. The temperature of the gas and the chemical potential $\mu_0$ in the center of the trap are independently extracted from fits of the averaged singles density profiles $\langle \tilde{s}_{\uparrow,\downarrow}(\mu) \rangle$ to NLCE data (see Methods). Due to the occurrence of unphysical oscillations in the NLCE data around quarter filling\cite{Khatami2011} our temperature determination is restricted to temperatures above $k_B T/t \approx 0.6$.

In Figure 1a we show the density profiles $\langle \tilde{s}_\uparrow(\bm{r}) \rangle$ and $\langle \tilde{s}_\downarrow(\bm{r}) \rangle$ for $U/t=8.2(5)$ and $k_BT/t = 0.96(2)$ averaged over a set of $130$ experimental runs. Figure 1b and 1c show the spatial distribution of the zero-momentum spin structure factor $S_{\bm{q}=0}(\bm{r})$ and the corresponding average value taken over regions of constant chemical potential $S_{\bm{q}=0}(\mu)$, respectively. In order to infer from our data  the presence of AFM correlations between different sites, we compare the spin structure factor with the local moment $C_{0,0} = \braket{(\hat{S}^z_{i})^2}-\langle \hat{S}^z_{i} \rangle^2$, which represents the on-site fluctuations of $\hat{S}^z$ and hence the local contribution to the magnetic susceptibility (see Figure 1c). Due to the Pauli principle, the local moment is obtained directly from the singles distributions according to $C_{0,0} =( \langle \hat{s}_\uparrow \rangle+\langle \hat{s}_\downarrow \rangle)/4$ (see Methods). While for low filling the structure factor closely follows the local moment, a clear mismatch between the structure factor and the local moment is observed around half-filling $\mu = U/2$, indicating the emergence of non-local AFM correlations.

We first focus on the situation at half-filling where for strong interactions the system enters a Mott-insulating state and particle-hole symmetry permits accurate theoretical simulations with which to compare our data.  We extract the spin structure factor at $\mu=U/2$ from a symmetric-peak fit to the measured spin structure factor $S_{\bm{q}=0}(\mu)$ (see Methods). In Figure 2a we present the structure factor at half-filling $S_{\bm{q}=0}(\mu=U/2)$ as a function of temperature for weak ($U/t=1.6$), intermediate ($U/t=8.2$), and strong ($U/t=12.0$) interactions. Within the uncertainties we find good agreement between our experimental data and numerical data from NLCE\cite{Khatami2011} and QMC\cite{Paiva2010} calculations of the Hubbard model at half-filling. For a quantitative analysis of beyond-nearest-neighbour AFM correlations at half-filling we also plot in Figure 2a NLCE data for $|C_{0,0}|-4 |C_{0,1}|$. For intermediate and strong interactions at temperatures $k_B T < t$ a comparison with the spin structure factor indicates that  contributions from next-nearest-neighbour AFM correlations $C_{1,1}$ become relevant and we observe these to be positive as expected. As is evident from Figure 2a, the spin structure factor can be used as a thermometer in the range $k_B T \lesssim t$ where the density distribution responds weakly upon further cooling\cite{Cocchi2016}, and our data constitute an initial calibration of this technique.

\begin{figure*}
 \includegraphics[width=2\columnwidth,clip=true]{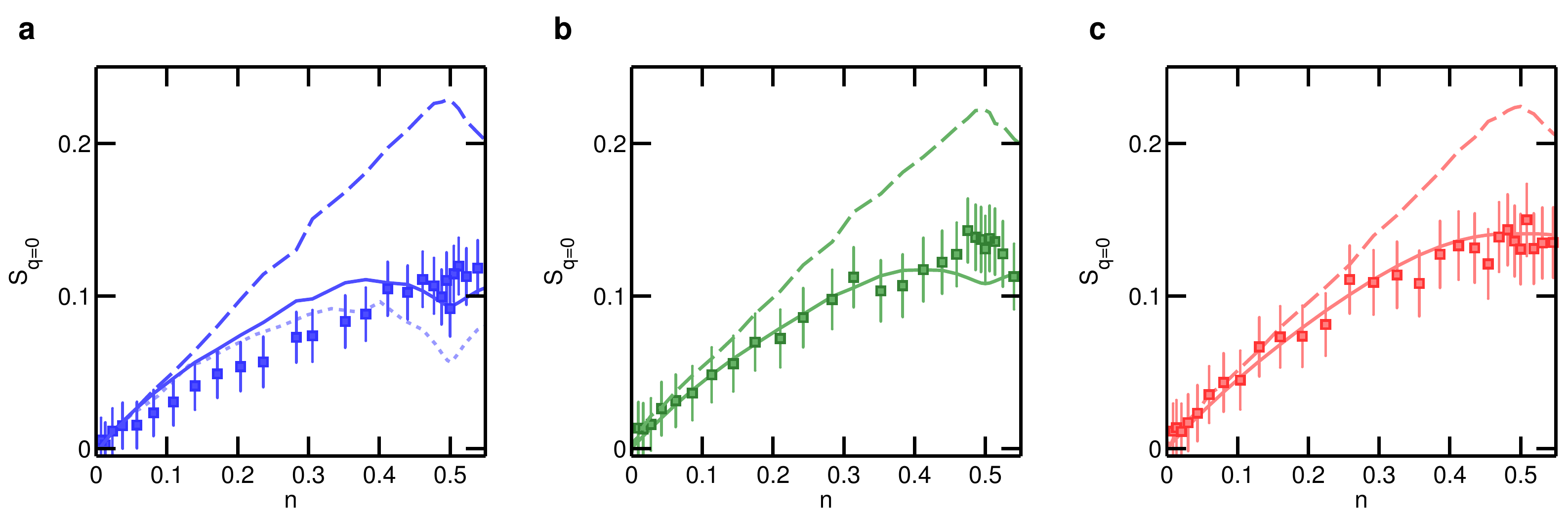}
 \caption{Effect of doping a Mott insulator. We compare $S_{\bm{q}=0}$ vs. filling $n$ at $U/t=8.2(5)$ for temperatures (a) $k_B T/t = 0.63(1)$, (b) $0.71(2)$ and (c) $0.96(2)$. The dashed lines show the measured local moment and the solid lines show NLCE calculations of local moment plus nearest-neighbour correlations, $|C_{0,0}|-4 |C_{0,1}|$. For all data a gradual disappearance of magnetic correlations with increasing doping $0.5-n$ is observed. The dotted line in (a) shows the NLCE result for a temperature of $k_BT/t=0.5$ showing better agreement with the data at low fillings.}
 \label{fig3}
\end{figure*}

Employing the fluctuation-dissipation theorem, we compute the magnetic susceptibility $\chi$ from the spin structure factor, which is shown in Figure 2b together with numerical data. Generally, we observe an increase of $\chi$ with interaction caused by the suppression of doubly-occupied sites which otherwise prevent the formation of a local moment.  

Quantum magnetism in the Hubbard model at half-filling and for strong interactions maps to the Heisenberg model of localized spins with the effective spin-spin interaction constant $J=4t^2/U$\cite{Anderson1959}. The physics of the Heisenberg model is governed by the single parameter $J$ and emerges for temperatures $k_B T\sim J$. In Figure 2c we show the agreement of the data for medium and strong interactions, $U/t>8$, with the prediction of the Heisenberg model when both temperature and magnetic susceptibility are scaled in units of $J$. As expected, we observe that for weak interactions, $U/t=1.6$, the Hubbard model does not map to the Heisenberg model due to large on-site density fluctuations which prevent the buildup of strong local moments.

Our system allows us to investigate the reduction of AFM spin correlations upon doping the Mott-insulator with holes at constant temperature. The ability to tune doping over an arbitrary range is impossible in the solid state, demonstrating the utility of cold atomic gases for quantum simulation of fundamental physics. In Figure 3 we show the measured spin structure factor for different fillings, $n = \langle\hat{n}_{i,\uparrow}+\hat{n}_{i,\downarrow}\rangle/2$, and temperatures. We compare the measured spin structure factor with the measured local moment (dashed line) and the theoretically expected local moment plus nearest-neighbour contributions $|C_{0,0}|-4 |C_{0,1}|$ (solid line) from NLCE calculations. For the two coldest temperatures, the measured structure factor at half-filling exceeds the sum of local moment (positive contribution) and nearest-neighbour correlations (negative contribution) indicating the presence of  next-nearest neighbour correlations, which have the same sign as the local moment. As the chemical potential is reduced and thereby the Mott insulator is doped with holes, we find that, at a filling  of $n \sim 0.4$ and temperatures $k_B T/t>0.7$, the spin correlations are fully described by nearest-neighbor AFM correlations and, further, at a filling below $n \sim 0.1$ only the local moment persists. Numerical simulations of the two-dimensional Hubbard model away from half-filling at low temperatures are very challenging and, while nearest-neighbour correlations have been computed\cite{Khatami2011}, theories of the magnetic susceptibility, which includes magnetic correlations at all length scales, have  been published only for very specific parameter sets\cite{Moreo1993,Bonca2003}. We also remark that even for low filling our coldest data in Figure 3a are systematically below the NLCE prediction for a temperature of $k_BT/t=0.63$, which is the value given by our density-based thermometry. Instead, the data is better described by NLCE and QMC for lower temperature  $k_BT/t = 0.5$ (see dotted line in Figure 3a and crosses in Figure 2a), which could indicate that the density-based thermometry overestimated the temperature in this measurement.

In the future, our in-situ imaging of the two-dimensional lattice could be extended by magnetic resonance techniques to imprint spin-waves\cite{Koschorreck2013,Hild2014} and detect the spin structure factor at arbitrary wave vectors. Such an addition could permit the measurements presented here to be complemented by studies of the dynamics of transport and diffusion in the spin sector of the two-dimensional Hubbard model. Furthermore, momentum-resolved Raman spectroscopy applied locally in the gas could help to identify pseudogap physics expected for finite doping by measuring dynamical correlation functions\cite{Dao2007,Stewart2008,Feld2011}.

\section{Methods}

\subsection{Spin-sensitive detection}
After preparation of the system with the desired parameters, we freeze the atomic motion by increasing the depth of the horizontal lattices to $60 E_\mathrm{rec}$ within $\unit[1]{ms}$. The horizontal lattice beams cross under an angle of $85.6(7)^\circ$ forming an almost square lattice structure. To remove atoms residing on doubly-occupied sites, we use adiabatic radio-frequency (RF) sweeps to transfer the spin-down population into the hyperfine state $|F=9/2,m_F=-3/2\rangle$. Spin-down atoms on doubly-occupied sites then undergo exothermic spin-changing collisions with the spin-up state $|F=9/2,m_F=-9/2\rangle$. Since the energy release from these collisions exceeds the trap depth by more than a factor of 10, atoms residing on doubly-occupied sites are lost from the trap. We experimentally confirmed that the spatial distribution of singly-occupied sites is unaffected by the removal of doubly-occupied sites.

Using high-resolution RF tomography in a vertical magnetic field gradient we transfer both spin components within a single horizontal layer simultaneously into unpopulated hyperfine states, which are imaged consecutively on the $|F=9/2, m_F=-9/2\rangle$ to $|F'=11/2, m_F'=-11/2\rangle$ cycling transition. Further details on the preparation of the system, the calibration of the lattice parameters and the imaging can be found in our previous work\cite{Cocchi2016}.

\subsection{Correlation analysis}
The recorded in-situ images $\{\tilde{s}_\sigma(\bm{r})\}_{\sigma = \uparrow, \downarrow}$ represent individual spatial distributions $s_{i,\sigma}$ of atoms in spin state $\sigma$ residing on singly-occupied lattice sites $i$ which are convolved with the imaging point spread function $p(\mathbf{r})$ according to $\tilde{s}_\sigma(\bm{r}) = \sum_i p(\bm{r}-\bm{r}_i) s_{i,\sigma}$ where $\int d^2\bm{r} p(\bm{r}) = 1$. The spatial correlation function of $\tilde{S}^z(\bm{r}) = (\tilde{s}_\uparrow(\bm{r})-\tilde{s}_\downarrow(\bm{r}))/2$ evaluated over the set of images at positions $\bm{r}$ and $\bm{r}'$ therefore contains contributions from the underlying spin-spin correlations between different lattice sites $i$ and $j$:
\begin{align}
C(\bm{r},\bm{r}') &= \langle \tilde{S}^z(\bm{r})\tilde{S}^z(\bm{r}')\rangle-\langle \tilde{S}^z(\bm{r})\rangle \langle\tilde{S}^z(\bm{r}')\rangle\notag\\ 
&= \sum_{i,j} p(\bm{r}-\bm{r}_i)p(\bm{r}'-\bm{r}_j)(\langle \hat{S}^z_i \hat{S}^z_j\rangle-\langle \hat{S}^z_i\rangle \langle \hat{S}^z_j\rangle).\notag
\end{align}
Integrating the correlation function $C(\bm{r},\bm{r}')$ over a region of radius $\xi$ large compared to the imaging resolution $\rho = \unit[1.2(1)]{\mu m}$ (FWHM of the auto-correlation peak) and the correlation length $\xi_0$ results in
\begin{align}
C_{\xi}(\bm{r}) &\equiv \int_{|\bm{r}'-\bm{r}|\leq\xi} d^2\bm{r}' C(\bm{r},\bm{r}')\notag \\ &\stackrel{\xi\gg \xi_0,\rho}{=} \sum_i p(\bm{r}-\bm{r}_i) \sum_{j} (\langle \hat{S}^z_i \hat{S}^z_j\rangle-\langle \hat{S}^z_i\rangle \langle \hat{S}^z_j\rangle).\notag
\end{align}
The spin structure factor $S_{\bm{q}=0}= \sum_{j} (\langle \hat{S}^z_i \hat{S}^z_j\rangle-\langle \hat{S}^z_i\rangle \langle \hat{S}^z_j\rangle)$ at zero wave vector $\bm{q}$ is then extracted by integrating $C_{\xi}(\bm{r})$ over regions of constant chemical potential (with binsize $\Delta \mu = h\times\unit[100]{Hz}$) and dividing by the number of lattice sites contained in those regions. In the data analysis, we chose the size $\xi$ of the integration region in order to cover most of the point spread function while minimizing the statistical noise from uncorrelated regions. For the data shown as filled circles in Figures 1 and 2 of the main text we used $\xi = \unit[1.7]{\mu m}$ and rescaled the integrated correlations (for every temperature and interaction value shown) to match the measured local moment plus NLCE data of the nearest-neighbor correlations\cite{Khatami2011}, $(|C_{0,0}|-4 |C_{0,1}|)$, up to a filling of $n=0.25$ where longer-ranged correlations are negligible over the investigated temperature range (see Figure 4). For all the other data (shown in squares in Figures 1,2 and 3 of the main text) we evaluated the spin structure factor from a larger data set using $\xi  =\unit[4]{\mu m}$ and rescaled the data by a common factor, which we extracted at the highest temperature of this data set, $k_B T/t = 0.96(2)$, in the same way as described above.

\begin{figure}
 \includegraphics[width=0.7\columnwidth]{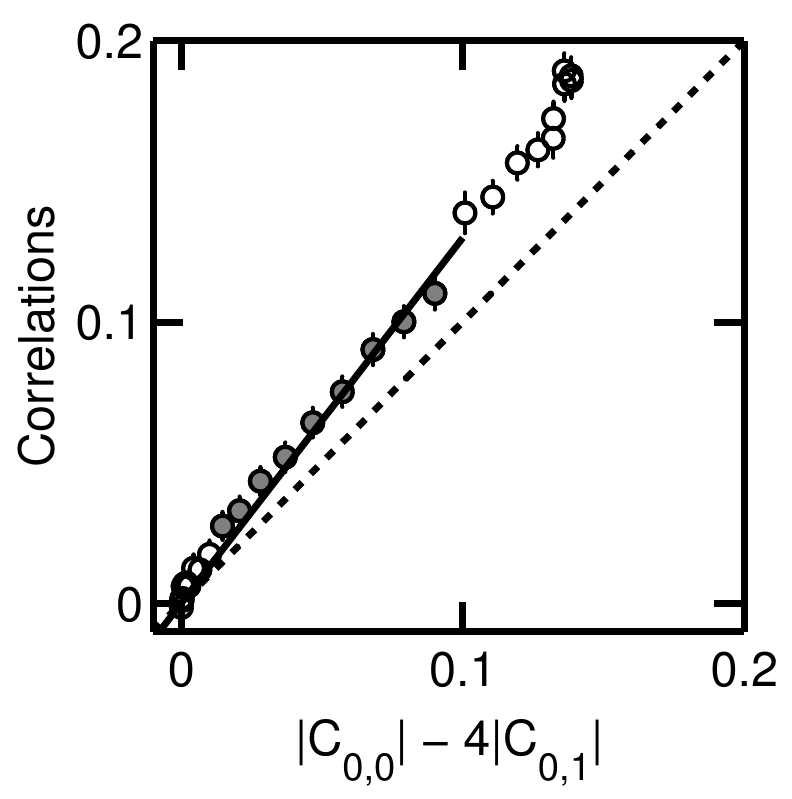}
 \caption{Rescaling of integrated correlations. Shown are the integrated correlations using $\xi = \unit[1.7]{\mu m}$ for an exemplary data set ($k_B T = 0.96 $ and $U/t = 8.2$) versus $|C_{0,0}|-4 |C_{0,1}|$ extracted from experimental data ($C_{0,0}$) and NLCE data ($C_{0,1}$). The solid line shows a fit to the filled data points (corresponding to fillings in the range $0.03 \leq n \leq 0.25$), the slope is used for rescaling the data to fall onto the bisecting line (dashed).}
 \label{ExtDataFig}
\end{figure}

For extracting the value of the spin structure factor at half-filling we fitted $S_{\bm{q}=0}-(|C_{0,0}|-4 |C_{0,1}|)$, evaluated as a function of chemical potential, with a Gaussian centered at $\mu=U/2$, and added the resulting peak amplitude to $|C_{0,0}|-4 |C_{0,1}|$.
 
Additional technical noise in the images originating from photon shot noise, camera read-out noise and far off-resonantly imaged atoms residing in the adjacent lattice layers was taken into account by applying an identical correlation analysis on a set of background images, which were recorded for a far off-resonant RF tomography pulse, and subtracting the corresponding correlation function from $C_\xi(\bm{r})$.

To extract the local moment $C_{0,0} = \braket{(\hat{n}_{i,\uparrow}-\hat{n}_{i,\downarrow})^2}/4$ from the density distributions of singly-occupied sites, we use the Pauli principle $\braket{\hat{n}_{i,\sigma}^2} = \braket{\hat{n}_{i,\sigma}}$ and the relation $\braket{\hat{n}_{i,\sigma}} = \braket{\hat{s}_{i,\sigma}}+\braket{\hat{n}_{i,\uparrow} \hat{n}_{i,\downarrow}}$.

\subsection{Thermometry}
We extracted the temperature of the gas prepared in the 2D lattice from fits of the measured singles distributions of spin-up and spin-down atoms to NLCE data\cite{Khatami2011}. For the intermediate and strong interaction data shown in the main text we used the calculated interaction strengths\cite{Cocchi2016} $U/t=8.2$ and $U/t=12$ and find excellent agreement between experimental data and NLCE fit. For the weak interaction data (with calculated interaction $U/t=1.6$), however, we are not able to fit the equilibrated singles and doubles distributions as well unless allowing the interaction parameter to go free which results in a fitted value of $(U/t)_\mathrm{fit} = 2.4$. We confirmed the resulting temperatures (which are considerably lower as compared to those resulting from the fit using the calculated value of $U/t=1.6$) by fitting the low-filling wings of the cloud to theory of the non-interacting Fermi gas\cite{Drewes2016}.

\begin{acknowledgments}
We thank C. Kollath for discussion and N. Wurz for experimental assistance. The work has been supported by DFG, the Alexander-von-Humboldt Stiftung, EPSRC and ERC.
\end{acknowledgments}

\end{document}